\documentclass[journal]{IEEEtran}

\ifCLASSINFOpdf
\else
\fi

\usepackage{setspace}
\usepackage{bbm}
\usepackage[cmex10]{amsmath}
\usepackage{amssymb}
\usepackage{cite}
\usepackage{graphicx}
\usepackage{array,color}
\usepackage{amsmath}
\allowdisplaybreaks
\usepackage{stfloats}
\usepackage{graphicx}
\usepackage{epstopdf}
\usepackage{subfigure}
\usepackage{tabularx}
\usepackage{epsfig,epsf,color,balance,cite}
\usepackage{algorithmic}
\usepackage{algorithm}
\usepackage{url}
\usepackage{bm}
\usepackage{multirow}

\usepackage{amsthm}

\ifodd 0
\usepackage{soul}
\usepackage{color}
\setstcolor{red}

\newcommand{\rev}[1]{{\color{red}#1}} 
\newcommand{\del}[1]{\st{#1}} 

\newcommand{\com}[1]{\textbf{\color{red} (COMMENT: #1)}} 
\newcommand{\response}[1]{\textbf{\color{green} (RESPONSE: #1)}} 
\else

\newcommand{\rev}[1]{#1}

\newcommand{\del}[1]{}

\newcommand{\com}[1]{}
\newcommand{\comg}[1]{}
\newcommand{\response}[1]{}
\fi

\title{\huge {IRS Meets Relaying: Joint Resource Allocation and Passive Beamforming Optimization}}

\author{ 
	Beixiong Zheng,~\IEEEmembership{Member,~IEEE}, and Rui Zhang,~\IEEEmembership{Fellow,~IEEE} 
	\vspace{-1.0cm}
	
	\thanks{
		
	This work was supported in part by NUS Research Grants R-261-518-005-720 and R-263-000-E86-112. \emph{(Corresponding author: Beixiong Zheng.)}
		
		The authors are with the Department of Electrical and Computer Engineering, National University of Singapore, Singapore 117583,
		email: \{elezbe, elezhang\}@nus.edu.sg.

	}
}

\begin{document}
\markboth{IEEE Wireless Communications Letters, Vol. XX, No. XX, XXX 2021}{SKM: My IEEE article}
\maketitle
\begin{abstract}
	Intelligent reflecting surface (IRS) has recently emerged as a new solution to enhance wireless communication performance via passive signal reflection.
In this letter, we unlock the potential of IRS controller in relaying information and propose a novel IRS-assisted communication system with both IRS passive reflection and active relaying. Specifically, we jointly optimize the time allocations for decode-and-forward (DF) relaying by the IRS controller and the IRS passive beamforming to maximize the achievable rate of the proposed system.
We also compare the rate performance of the proposed system with the conventional IRS without relaying, and reveal the conditions for one to outperform the other. Simulation results demonstrate that the proposed new design can significantly improve the coverage/rate performance of IRS-assisted systems.
\end{abstract}
\begin{IEEEkeywords}
	Intelligent reflecting surface (IRS), decode-and-forward (DF) relay, resource allocation, passive beamforming.
\end{IEEEkeywords}
\IEEEpeerreviewmaketitle

\vspace{-0.6cm}
\section{Introduction}
\IEEEPARstart{I}{ntelligent} reflecting surface (IRS) has recently emerged as an innovative technology to reconfigure the wireless propagation environment via tunable passive signal reflection. Specifically, IRS consists of a large number of passive
reflecting elements that can be tuned by an IRS controller to alter the phases and/or amplitudes of their reflected signals with ultra-low power consumption, for facilitating the wireless transmissions in various applications and system setups (see e.g. \cite{wu2020intelligent,Renzo2020Smart} and the references therein).

In the existing works on IRS, IRS is usually deployed near its assisted wireless devices to enhance the communication performance in terms of rate or range from their associated base station (BS)/access point (AP).
This is due to the passive signal reflection of IRS without signal amplification/regeneration as well as the severe ``product-distance" path-loss of the IRS reflected channel, which result in the limited serving range for IRSs in practice, as compared to conventional active relays.
As such, performance comparison between the ``passive" IRS and ``active" relay was studied in \cite{Emill2020Intelligent,Renzo2020Reconfigurable,Ye2021Spatially} from different aspects.
Recently, active relay was also added to the IRS-assisted communication system for coverage extension \cite{Yildirim2021Hybrid,ying2020relay,nguyen2021hybrid}, which, however, inevitably incurs higher cost and complexity for implementation.

\begin{figure}[!t]
	\centering
	\includegraphics[width=2.5in]{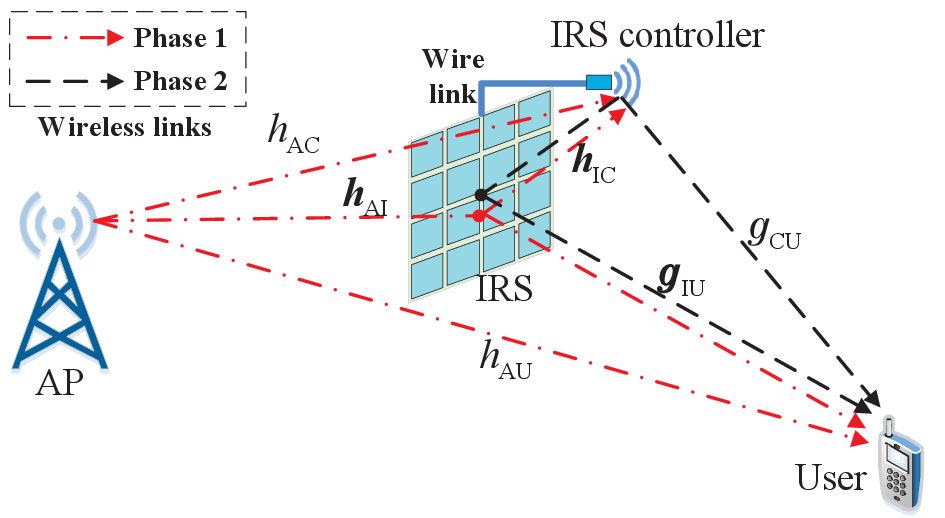}
	\setlength{\abovecaptionskip}{-3pt}
	\caption{\rev{A relaying IRS-assisted communication system.}}
	\label{system}
	\vspace{-0.3cm}
\end{figure}

To overcome the limited IRS coverage issue without increasing the deployment/hardware cost, in this letter we propose a novel IRS-assisted communication system by unlocking the IRS controller for relaying information in addition to its conventional role of tuning the reflections of IRS elements only, as shown in Fig.~\ref{system}.
By enabling the IRS controller to actively relay the information from the AP to the user, the IRS can potentially help improve the transmission rate even if the user is out of the range of IRS signal reflection, thus greatly enhancing the coverage performance of conventional IRSs with passive signal reflection only.  First, we formulate the joint time allocation and passive beamforming optimization problem for the proposed system with both IRS reflection and controller relaying to maximize the achievable rate of its assisted user.
Then, by analytically comparing the rate performance of the proposed system with the conventional IRS without controller relaying, we unveil the sufficient conditions for one to perform better than the other. Next, we propose an efficient algorithm based on the alternating optimization (AO) technique to solve the formulated problem sub-optimally. Finally, we present simulation results to show the practical advantages of the proposed system under a practical setup.

\emph{Notation:} 
For a complex-valued vector $\bm{x}$, $\| \bm{x} \|_1$ denotes its $\ell_1 $-norm and $\angle (\bm{x} )$ returns the phase of each element in $\bm{x}$.

\vspace{-0.5cm}
\section{System Model and Transmission Protocol}\label{sys}

As shown in Fig.~\ref{system}, we consider an IRS-assisted downlink communication system, where the transmission from an AP to its associated user is aided by an IRS that consists of $M$ passive reflecting elements \rev{connected to a smart controller via a wire link.}
Note that besides adjusting the IRS reflection and exchanging (control/channel) information with the AP, the IRS controller also acts as a potential decode-and-forward (DF) relay in our considered system to enhance the user rate performance and/or IRS coverage range. 
Thus, we refer to this new IRS-assisted communication scenario with IRS controller relaying as the ``relaying IRS-assisted communication".  

In this letter, we consider a two-phase orthogonal-time transmission protocol for the relaying IRS-assisted communication, where the AP transmits its message to both the user and the IRS controller (or relay) in Phase~1; and the IRS controller (relay) forwards the message to the user in Phase~2 if it can decode the message successfully.\footnote{Without causing confusion, we use ``IRS controller" and ``relay" interchangeably in this letter.} Moreover, given the total transmission time, we let $\alpha$ and $1-\alpha$ denote the time fractions allocated to Phases~1 and 2, respectively, where $0\le \alpha \le 1$.
For the purpose of exposition, we assume that the AP, user, and IRS controller are all equipped with one single antenna\footnote{\rev{For the case of multi-antenna AP and/or relay, this work needs to be extended to consider joint active and passive beamforming design.}}. 
Let $h_{\rm AU}\in {\mathbb{C}}$, ${\bm h}_{\rm AI}\in {\mathbb{C}^{M\times 1 }}$, $h_{\rm AC}\in {\mathbb{C}}$, ${\bm h}_{\rm IC}\in {\mathbb{C}^{M\times 1 }}$, ${\bm g}_{\rm IU}\in {\mathbb{C}^{M\times 1 }}$, and $g_{\rm CU}\in {\mathbb{C}}$
denote the baseband equivalent channels for the AP$\rightarrow$user, AP$\rightarrow$IRS, AP$\rightarrow$controller, IRS$\rightarrow$controller, IRS$\rightarrow$user, and controller$\rightarrow$user links, respectively.
As the first work on characterizing the fundamental performance gain brought by the IRS controller relaying, we assume for simplicity that
the perfect channel state information (CSI) of all channels involved is available at the AP/IRS controller and the channel reciprocity holds for each link.
In general, the IRS can change its reflection coefficients over the two phases.
As such, we let ${\bm \theta}_\mu\triangleq[{\theta_{\mu,1}},{\theta_{\mu,2}},\ldots,{\theta_{\mu,M}}]^T\in {\mathbb{C}^{M\times 1 }}$ denote the equivalent reflection vector of IRS in Phase $\mu$, where the reflection amplitudes of all elements are set to one or the maximum value to maximize the signal reflection power and thus $|{\theta_{\mu,m}}|=1, \forall m=1,\ldots,M$, $\mu\in \{1,2\}$.

More specifically, in Phase 1, the AP transmits $\sqrt{P_A} s_A$ to the user and the IRS controller simultaneously, where $s_A\in {\mathbb{C}}$ denotes the information symbol with ${\mathbb E}\{|s_A|^2\}=1$, \rev{and $P_A$ is the (peak) transmit power of the AP.}
Thus, the received signals at the user and the IRS controller during Phase 1 are respectively expressed as
\vspace{-0.2cm}
\begin{align}
y_U&=\left(h_{\rm AU}+{\bm h}_{\rm AI}^T {\bm \Phi}_1 {\bm g}_{\rm IU}\right)\sqrt{P_A} s_A +n_U\notag\\
&= \left(h_{\rm AU}+{\bm h}_{\rm AI}^T \text{diag} \left( {\bm g}_{\rm IU} \right){\bm \theta}_1 \right)\sqrt{P_A} s_A +n_U,\label{rec_user1}\\
y_C&=\left(h_{\rm AC}+{\bm h}_{\rm AI}^T {\bm \Phi}_1 {\bm h}_{\rm IC}\right)\sqrt{P_A} s_A +n_C\notag\\
& =\left(h_{\rm AC}+{\bm h}_{\rm AI}^T \text{diag} \left( {\bm h}_{\rm IC} \right) {\bm \theta}_1\right)\sqrt{P_A} s_A +n_C,\label{rec_con}
\end{align}
where $n_U$ and $n_C$ denote the zero-mean additive white Gaussian noise (AWGN) with variance $\sigma^2$ received at the user and the IRS controller, respectively, and
${\bm \Phi}_1=\text{diag} \left( {\bm \theta}_1 \right)$ denotes the diagonal reflection matrix of the IRS in Phase 1.
For notational convenience, we let ${\bm q}_U^H\triangleq {\bm h}_{\rm AI}^T \text{diag} \left( {\bm g}_{\rm IU} \right)$
and ${\bm q}_C^H\triangleq {\bm h}_{\rm AI}^T \text{diag} \left( {\bm h}_{\rm IC} \right)$ denote the cascaded AP$\rightarrow$IRS$\rightarrow$user channel and AP$\rightarrow$IRS$\rightarrow$controller channel, respectively.
Based on \eqref{rec_user1} and \eqref{rec_con}, the received signal-to-noise ratios (SNRs)  at the user and the IRS controller from the AP in Phase 1 are given by
\vspace{-0.2cm}
\begin{align}
&\rho_U \left( {\bm \theta}_1 \right)\triangleq \frac {P_A|h_{\rm AU}+{\bm q}_U^H{\bm \theta}_1|^2}{\sigma^2},\label{SNR_U1}\\
&\rho_C \left( {\bm \theta}_1 \right)\triangleq \frac {P_A|h_{\rm AC}+{\bm q}_C^H {\bm \theta}_1|^2}{\sigma^2},\label{SNR_C1}
\end{align}
respectively, which are both functions of ${\bm \theta}_1$.

If the IRS controller successfully decodes the message from the AP in Phase~1, it first re-encodes the message using incremental coding \cite{Liang2005Gaussian} and then transmits $\sqrt{P_C} s_C$ to the user during Phase 2, with the corresponding received signal at the user given by
\vspace{-0.2cm}
\begin{align}
{\tilde y}_U&=\left(h_{\rm CU}+{\bm h}_{\rm IC}^T {\bm \Phi}_2 {\bm g}_{\rm IU}\right)\sqrt{P_C} s_C +n_U\notag\\
&= \left(h_{\rm CU}+{\bm h}_{\rm IC}^T \text{diag} \left( {\bm g}_{\rm IU} \right){\bm \theta}_2 \right)\sqrt{P_C} s_C +{\tilde n}_U,\label{rec_user2}
\end{align}
where $s_C\in {\mathbb{C}}$ denotes the information symbol with ${\mathbb E}\{|s_C|^2\}=1$, \rev{$P_C$ is the (peak) transmit power of the IRS controller,}
${\tilde n}_U$ is the zero-mean AWGN with variance $\sigma^2$ during Phase 2,
and ${\bm \Phi}_2=\text{diag} \left( {\bm \theta}_2 \right)$ denotes the diagonal reflection matrix of the IRS in Phase~2.
Let ${\tilde{\bm q}}_U^H\triangleq{\bm h}_{\rm IC}^T \text{diag} \left( {\bm g}_{\rm IU} \right)$ denote the cascaded controller$\rightarrow$IRS$\rightarrow$user channel, and thus the received SNR at the user from the IRS controller in Phase~2 is given by
\vspace{-0.2cm}
\begin{align}
&{\tilde \rho}_U \left( {\bm \theta}_2 \right)\triangleq \frac {P_C|h_{\rm CU}+{\tilde{\bm q}}_U^H{\bm \theta}_2|^2}{\sigma^2},\label{SNR_U2}
\end{align}
which is a function of ${\bm \theta}_2$.
\vspace{-0.3cm}
\section{Problem Formulation and Rate Analysis}
\subsection{Problem Formulation}
For the relaying IRS-assisted communication system, the achievable rate in bits per second per Hertz (bps/Hz) based on the Gaussian orthogonal relay channel \cite{Liang2005Gaussian} is given by 
\vspace{-0.2cm}
\begin{align}\label{rate2}
&C_1\left({\bm \theta}_1,{\bm \theta}_2,\alpha\right)
={\text{min}} \Big\{\alpha \underbrace{\log_2\left(1+\rho_U \left( {\bm \theta}_1 \right)\right)}_{R_U \left( {\bm \theta}_1 \right)}\notag\\
&+(1-\alpha)\underbrace{\log_2\left(1+{\tilde \rho}_U \left( {\bm \theta}_2 \right)\right)}_{{\tilde R}_U \left( {\bm \theta}_2 \right)},\alpha \underbrace{\log_2\left(1+\rho_C \left( {\bm \theta}_1 \right)\right)}_{R_C \left( {\bm \theta}_1 \right)}\Big\}.
\end{align}
where $R_U \left( {\bm \theta}_1 \right)$, ${\tilde R}_U \left( {\bm \theta}_2 \right)$, and $R_C \left( {\bm \theta}_1 \right)$
denote the rates associated with the SNRs $\rho_U \left( {\bm \theta}_1 \right)$, ${\tilde \rho}_U \left( {\bm \theta}_2 \right)$,
and $\rho_C \left( {\bm \theta}_1 \right)$, respectively.
Note that the achievable rate in \eqref{rate2} holds only when the effective channel gain from the AP to the IRS controller is larger than that from the AP to the user, i.e., $\rho_C \left( {\bm \theta}_1 \right)> \rho_U \left( {\bm \theta}_1 \right)$ \cite{Liang2005Gaussian}. For convenience, we refer to this case as the ``relaying IRS case'' since the IRS controller relaying is used in addition to the IRS passive reflection.  
Also note that ${\bm \theta}_2$ should be set to maximize ${\tilde \rho}_U \left( {\bm \theta}_2 \right)$ in \eqref{rate2} so as to maximize the achievable rate $C_1\left({\bm \theta}_1,{\bm \theta}_2,\alpha\right)$, i.e., ${\tilde \rho}_U^{\star}=\underset{{\bm \theta}_2}{\text{max}}~~  {\tilde \rho}_U \left( {\bm \theta}_2 \right)$. Accordingly, the optimal passive reflection vector during Phase~2 (denoted by ${\bm \theta}_{2}^{\star}$) is designed as
\vspace{-0.2cm}
\begin{align}\label{opt_ph2}
{\bm \theta}_{2}^{\star}=\arg \underset{{\bm \theta}_2}{\text{max}}~~  {\tilde \rho}_U \left( {\bm \theta}_2 \right)=e^{j\angle \left(h_{\rm CU}\right)+\angle\left({\tilde{\bm q}}_U\right)},
\end{align}
which achieves the maximum SNR as
${\tilde \rho}_U^{\star}=\frac {P_C\left(|h_{\rm CU}|+\|{\tilde{\bm q}}_U\|_1\right)^2}{\sigma^2}$ and the corresponding maximum rate as ${\tilde R}_U^{\star}=\underset{{\bm \theta}_2}{\text{max}}~{\tilde R}_U \left( {\bm \theta}_2 \right)=
\log_2\left(1+{\tilde \rho}_U^{\star}\right)$.
By substituting \eqref{opt_ph2} into \eqref{rate2}, the achievable rate can be simplified as
\vspace{-0.2cm}
\begin{align}\label{rate_relay}
&C_1\left({\bm \theta}_1,\alpha\right) \triangleq\underset{{\bm \theta}_2}{\text{max}}~~  C_1\left({\bm \theta}_1,{\bm \theta}_2,\alpha\right)\notag\\
&={\text{min}} \Big\{\alpha R_U\left( {\bm \theta}_1 \right) +(1-\alpha){\tilde R}_U^{\star}, \alpha R_C\left( {\bm \theta}_1 \right)\Big\}.
\end{align}
Next, we aim to maximize the achievable rate in \eqref{rate_relay} by jointly optimizing the passive reflection vector ${\bm \theta}_1$ and the time allocation factor $\alpha$, which can be formulated as
\vspace{-0.2cm}
\begin{align}
\text{(P1):}~
C_1^{\star}\triangleq& \underset{{\bm \theta}_1,\alpha}{\text{max}} 
& &  C_1\left({\bm \theta}_1,\alpha\right) \label{obj_P1}\\
& \text{~~s.t.} & & \rho_C \left( {\bm \theta}_1 \right)> \rho_U \left( {\bm \theta}_1 \right),\label{con0_P1}\\
& & & 0\le \alpha \le 1,\\ 
& & & |{\theta_{1,m}}|=1, \forall m=1,\ldots,M\label{con2_P1}.
\end{align}

However, if the relaying IRS case does not hold, i.e., $\rho_C \left( {\bm \theta}_1 \right)\le \rho_U \left( {\bm \theta}_1 \right)$, the IRS controller should not decode the message from the AP and the IRS only needs to reflect the signal from the AP to the user in Phase~1 (i.e., $\alpha=1$ or Phase 2 is not used). In this case,
our considered system reduces to the conventional IRS-assisted communication without controller relaying, for which the achievable rate is given by
\vspace{-0.2cm}
\begin{align}\label{rate_conv}
	C_2\left({\bm \theta}_1\right)=R_U \left( {\bm \theta}_1 \right)=
	\log_2\left(1+\rho_U \left( {\bm \theta}_1 \right)\right).
\end{align} 
This case is thus referred to as the ``conventional IRS case" since only the IRS passive reflection is employed.
It can be verified that the achievable rate maximization in \eqref{rate_conv} is equivalent to the SNR maximization in \eqref{SNR_U1}, i.e., finding $\rho_U^{\star}=\underset{{\bm \theta}_1}{\text{max}}~~  \rho_U \left( {\bm \theta}_1 \right)$; thus, the optimal passive reflection vector (denoted by ${\bm \theta}_{1,U}^{\star}$) is designed as
\vspace{-0.2cm}
\begin{align}\label{opt_vec}
{\bm \theta}_{1,U}^{\star}=\arg \underset{{\bm \theta}_1}{\text{max}}~~  \rho_U \left( {\bm \theta}_1 \right)=e^{j\angle \left(h_{\rm AU}\right)+\angle\left({\bm q}_U\right)},
\end{align}
which achieves the maximum SNR as $\rho_U^{\star}=\frac {P_A\left(|h_{\rm AU}|+\|{\bm q}_U\|_1\right)^2}{\sigma^2}$ and the corresponding maximum achievable rate in the conventional IRS case as 
\vspace{-0.2cm}
\begin{align}\label{opt_C2}
\hspace{-0.2cm}C_2^{\star}\triangleq \underset{{\bm \theta}_1}{\text{max}}~~  C_2\left({\bm \theta}_1\right)=\underset{{\bm \theta}_1}{\text{max}}~~  R_U\left({\bm \theta}_1\right)=\log_2\left(1+\rho_U^{\star}\right).
\end{align}

As compared to the conventional IRS case, it is more challenging to solve (P1) for the relaying IRS case\footnote{\rev{For the relaying IRS case, one practical channel estimation approach is to estimate the cascaded/direct CSI $\left\{{\bm q}_C^H, h_{\rm AC}\right\}$ and $\left\{{\tilde{\bm q}}_U^H, h_{\rm CU}\right\}$ at the AP and user, respectively, based on the pilot signals sent from the IRS controller \cite{zheng2019intelligent}. While for the conventional IRS without relaying, existing cascaded channel estimation methods (e.g., \cite{zheng2019intelligent}) can be applied to acquire the cascaded/direct CSI $\left\{ {\bm q}_U^H, h_{\rm AU}\right\}$.}}, which will be addressed in Section~\ref{Solution}.
\vspace{-0.3cm}
\subsection{Rate Analysis}
In this subsection, we first compare the maximum achievable rates $C_1^{\star}$ and $C_2^{\star}$ for the relaying IRS and conventional IRS cases, in order to reveal the practical conditions under which one case performs better than the other, or vice versa.  

First, let ${\bm \theta}_{1,C}^{\star}$ denote the passive reflection vector that maximizes the received SNR in \eqref{SNR_C1} at the IRS controller from the AP in Phase 1, which is given by
\begin{align}\label{opt_vec_con}
{\bm \theta}_{1,C}^{\star}=\arg \underset{{\bm \theta}_1}{\text{max}}~~  \rho_C \left( {\bm \theta}_1 \right)=e^{j\angle \left(h_{\rm AC}\right)+\angle\left({\bm q}_C \right)},
\end{align}
and the resultant maximum SNR is $ \rho_C^{\star}=\frac {P_A\left(|h_{\rm AC}|+\|{\bm q}_C\|_1\right)^2}{\sigma^2}$ and the corresponding maximum rate is ${ R}_C^{\star}=\underset{{\bm \theta}_1}{\text{max}}~R_C \left( {\bm \theta}_1 \right)=\log_2\left(1+\rho_C^{\star}\right)$.
Then, we have the following proposition.

\indent\emph{Proposition 1}: If ${\text{min}}\left\{{\tilde \rho}_U^{\star},\rho_C^{\star}\right\} \le \rho_U^{\star}$, then we have 
$C_1^{\star}\le C_2^{\star}$, i.e., the maximum achievable rate in the relaying IRS case is no larger than that in the conventional IRS case.
\begin{IEEEproof}
	We let $\Delta\left({\bm \theta}_1,\alpha\right)\triangleq C_1\left({\bm \theta}_1,\alpha\right)-C_2^{\star}$ denote the gap between the achievable rate of the relaying IRS case in \eqref{rate_relay} and the maximum achievable rate of the conventional IRS case in \eqref{opt_C2}, which can be further expressed as
	\begin{align}\label{gap}
	&\Delta\left({\bm \theta}_1,\alpha\right)
	\hspace{-0.1cm}={\text{min}} \Big\{\hspace{-0.1cm}\alpha R_U \hspace{-0.1cm}\left( {\bm \theta}_1 \right)\hspace{-0.06cm}+\hspace{-0.06cm}(1\hspace{-0.06cm}-\hspace{-0.06cm}\alpha){\tilde R}_U^{\star}-C_2^{\star}, \alpha R_C \hspace{-0.06cm}\left( {\bm \theta}_1 \right)\hspace{-0.06cm}-\hspace{-0.06cm}C_2^{\star}\hspace{-0.1cm}\Big\}\notag\\
	&={\text{min}} \Big\{ \underbrace{\alpha \left( R_U \left( {\bm \theta}_1 \right) -C_2^{\star}\right) + (1-\alpha)\left({\tilde R}_U^{\star}-C_2^{\star}\right)}_{\Delta_1\left({\bm \theta}_1,\alpha\right)},\notag\\ 
	&\qquad\qquad \underbrace{\alpha \left(R_C \left( {\bm \theta}_1 \right)-C_2^{\star}\right)-(1-\alpha)C_2^{\star}}_{\Delta_2\left({\bm \theta}_1,\alpha\right)}\Big\}.
	\end{align}
	If ${\text{min}}\left\{{\tilde \rho}_U^{\star},\rho_C^{\star}\right\} \le \rho_U^{\star}$, we have either ${\tilde \rho}_U^{\star}\le \rho_U^{\star}$ or $\rho_C^{\star}\le \rho_U^{\star}$, leading to the following two cases.
	\begin{itemize}
	\item  For the case of ${\tilde \rho}_U^{\star}\le \rho_U^{\star}$, we have ${\tilde R}_U^{\star}\le C_2^{\star}$ and thus $(1-\alpha)\left({\tilde R}_U^{\star}-C_2^{\star}\right)\le 0$ with $0\le 1- \alpha \le 1$.
	Furthermore, since we have $R_U \left( {\bm \theta}_1 \right) -C_2^{\star}\le 0, \forall {\bm \theta}_1$ according to \eqref{opt_C2}, it can be readily verified that $\Delta_1\left({\bm \theta}_1,\alpha\right)\le 0, \forall \alpha, {\bm \theta}_1$ if ${\tilde \rho}_U^{\star}\le \rho_U^{\star}$.
	\item For the case of $\rho_C^{\star}\le \rho_U^{\star}$, we have $R_C \left( {\bm \theta}_1 \right)\le { R}_C^{\star}\le C_2^{\star}$ and thus $\alpha \left(R_C \left( {\bm \theta}_1 \right)-C_2^{\star}\right)\le 0, \forall \alpha, {\bm \theta}_1$.
	Since $-(1-\alpha)C_2^{\star}\le 0$, we can further obtain $\Delta_2\left({\bm \theta}_1,\alpha\right)\le 0, \forall \alpha, {\bm \theta}_1$ if $\rho_C^{\star}\le \rho_U^{\star}$.
	\end{itemize}
Moreover, according to \eqref{gap}, we can obtain that $\Delta\left({\bm \theta}_1,\alpha\right)\le 0, \forall \alpha, {\bm \theta}_1$ holds if and only if either $\Delta_1\left({\bm \theta}_1,\alpha\right)\le 0, \forall \alpha, {\bm \theta}_1$ or $\Delta_2\left({\bm \theta}_1,\alpha\right)\le 0, \forall \alpha, {\bm \theta}_1$. Based on the above, it follows that if ${\text{min}}\left\{{\tilde \rho}_U^{\star},\rho_C^{\star}\right\} \le \rho_U^{\star}$, we have $\Delta\left({\bm \theta}_1,\alpha\right)\le 0, \forall \alpha, {\bm \theta}_1$ and thus $C_1\left({\bm \theta}_1,\alpha\right)\le C_2^{\star}, \forall \alpha, {\bm \theta}_1$, which leads to $C_1^{\star}\triangleq \underset{{\bm \theta}_1,\alpha}{\text{max}} ~~C_1\left({\bm \theta}_1,\alpha\right)\le C_2^{\star}$, thus completing the proof.
%
\end{IEEEproof}

According to Proposition~1, ${\text{min}}\left\{{\tilde \rho}_U^{\star},\rho_C^{\star}\right\} \le \rho_U^{\star}$ is a sufficient condition for $C_1^{\star}\le C_2^{\star}$, which implies that the IRS controller relaying needs not to be employed in this case.  
 Furthermore, to show the condition under which the relaying IRS case is superior to the conventional IRS case in terms of maximum achievable rate, we have another proposition as follows.
 
 \indent\emph{Proposition 2}: If ${\tilde \rho}_U^{\star} > \rho_U^{\star}$ and there exists a ${\bm \theta}_1$ such that $R_C \left( {\bm \theta}_1 \right)>\frac{{\tilde R}_U^{\star}-R_U \left( {\bm \theta}_1 \right)}{{\tilde R}_U^{\star}-C_2^{\star}}C_2^{\star}$, then we have 
 $C_1^{\star}> C_2^{\star}$, i.e., the maximum achievable rate in the relaying IRS case is always larger than that in the conventional IRS case.
 \begin{IEEEproof}
 	The rate gap in \eqref{gap} can be rewritten as
 	\begin{align}\label{gap2}
 	\Delta\left({\bm \theta}_1,\alpha\right)={\text{min}} &\Big\{ \underbrace{\alpha \left( R_U \left( {\bm \theta}_1 \right) -{\tilde R}_U^{\star}\right) +  {\tilde R}_U^{\star}-C_2^{\star}}_{\Delta_1\left({\bm \theta}_1,\alpha\right)}, \notag\\
 	&\underbrace{\alpha R_C \left( {\bm \theta}_1 \right)-C_2^{\star}}_{\Delta_2\left({\bm \theta}_1,\alpha\right)}\Big\}.
 	\end{align}
 	If ${\tilde \rho}_U^{\star} > \rho_U^{\star}$, then we have ${\tilde R}_U^{\star}> C_2^{\star}\ge R_U \left( {\bm \theta}_1 \right)$ according to \eqref{opt_C2} and thus ${\tilde R}_U^{\star}- C_2^{\star}>0$ and $R_U \left( {\bm \theta}_1 \right) -{\tilde R}_U^{\star}<0$. In this case, to ensure $\Delta_1\left({\bm \theta}_1,\alpha\right)>0$, $\alpha$ should satisfy $\alpha<\frac{{\tilde R}_U^{\star}-C_2^{\star}}{{\tilde R}_U^{\star}-R_U \left( {\bm \theta}_1 \right)}$, where we have $\frac{{\tilde R}_U^{\star}-C_2^{\star}}{{\tilde R}_U^{\star}-R_U \left( {\bm \theta}_1 \right)}\le 1$ due to the fact that $0<{\tilde R}_U^{\star}-C_2^{\star}\le {\tilde R}_U^{\star}-R_U \left( {\bm \theta}_1 \right)$.
 	On the other hand, to ensure $\Delta_2\left({\bm \theta}_1,\alpha\right)>0$, $\alpha$ should satisfy $\alpha> \frac{C_2^{\star}}{R_C \left( {\bm \theta}_1 \right)}$. 
 	Accordingly, to ensure the existence of $\alpha$ such that both $\Delta_1\left({\bm \theta}_1,\alpha\right)>0$ and $\Delta_2\left({\bm \theta}_1,\alpha\right)>0$ can hold at the same time to achieve $\Delta\left({\bm \theta}_1,\alpha\right)>0$, we should have 
 	$\frac{{\tilde R}_U^{\star}-C_2^{\star}}{{\tilde R}_U^{\star}-R_U \left( {\bm \theta}_1 \right)}>\frac{C_2^{\star}}{R_C \left( {\bm \theta}_1 \right)}$, which is equivalent to $R_C \left( {\bm \theta}_1 \right)>\frac{{\tilde R}_U^{\star}-R_U \left( {\bm \theta}_1 \right)}{{\tilde R}_U^{\star}-C_2^{\star}}C_2^{\star}\ge C_2^{\star}$ due to $\frac{{\tilde R}_U^{\star}-R_U \left( {\bm \theta}_1 \right)}{{\tilde R}_U^{\star}-C_2^{\star}}\ge 1$.
 	
 	Based on the above, if ${\tilde \rho}_U^{\star} > \rho_U^{\star}$ and there exists a ${\bm \theta}_1$ such that  $R_C \left( {\bm \theta}_1 \right)>\frac{{\tilde R}_U^{\star}-R_U \left( {\bm \theta}_1 \right)}{{\tilde R}_U^{\star}-C_2^{\star}}C_2^{\star}$, then we can achieve $\Delta\left({\bm \theta}_1,\alpha\right)>0$ with $\frac{C_2^{\star}}{R_C \left( {\bm \theta}_1 \right)}<\alpha<\frac{{\tilde R}_U^{\star}-C_2^{\star}}{{\tilde R}_U^{\star}-R_U \left( {\bm \theta}_1 \right)}$, which leads to $C_1^{\star}\ge C_1\left({\bm \theta}_1,\alpha\right)>C_2^{\star}$, thus completing the proof.
\end{IEEEproof}
\section{Proposed Solution to (P1)}\label{Solution} 
In this section, we address how to solve (P1) for the relaying IRS case. Note that in (P1), based on \eqref{rate_relay}, we need to jointly design ${\bm \theta}_1$ and $\alpha$ to balance the achievable rates at the user (over two phases) and IRS controller (in Phase 1).
However, since the objective function in \eqref{obj_P1} is non-concave and the unit-modulus constraint in \eqref{con2_P1} is non-convex,
(P1) is generally difficult to be solved optimally. 
In the following, we solve (P1) sub-optimally by applying the AO technique, which alternately optimizes the
time allocation factor $\alpha$ and the passive reflection vector ${\bm \theta}_1$ during Phase 1 in an iterative manner, until
the convergence is achieved.

\subsubsection{\bf Time Allocation Optimization}
For any fixed feasible ${\bm \theta}_1$, (P1) reduces to the following time allocation problem.
\begin{align}
\hspace{-0.3cm}\text{(P2):}
C_1\hspace{-0.08cm}\left(\hspace{-0.05cm}{\bm \theta}_1\hspace{-0.05cm}\right) \hspace{-0.08cm}\triangleq\hspace{-0.12cm} \underset{0\le \alpha \le 1}{\text{max}} 
  {\text{min}} \Big\{\hspace{-0.08cm}\alpha R_U\hspace{-0.08cm}\left(\hspace{-0.05cm} {\bm \theta}_1 \hspace{-0.05cm}\right) \hspace{-0.08cm}+\hspace{-0.08cm}(1\hspace{-0.05cm}-\hspace{-0.05cm}\alpha){\tilde R}_U^{\star}, \alpha R_C\hspace{-0.05cm}\left(\hspace{-0.05cm} {\bm \theta}_1 \hspace{-0.05cm}\right)\hspace{-0.08cm}\Big\},\hspace{-0.1cm} \label{obj_P2}
\end{align}
which is a linear optimization problem over $\alpha$ and can be solved optimally, as given in the following proposition.

\indent\emph{Proposition 3}: The optimal $\alpha^{\star}$ to (P2) is 
\begin{align}\label{opt_factor}
\alpha^{\star}=\frac{{\tilde R}_U^{\star} }{ R_C\left( {\bm \theta}_1 \right)+{\tilde R}_U^{\star}-R_U\left( {\bm \theta}_1 \right)}.
\end{align}
\begin{IEEEproof}
Given feasible ${\bm \theta}_1$, (P2) is solved only under $\rho_C \left( {\bm \theta}_1 \right)> \rho_U \left( {\bm \theta}_1 \right)$ according to \eqref{con0_P1} and thus we have $R_C\left( {\bm \theta}_1 \right)> R_U\left( {\bm \theta}_1 \right)$. Moreover, to make problem \eqref{obj_P2} valid, we need to have ${\tilde R}_U^{\star}>C_2^{\star}\ge R_U\left( {\bm \theta}_1 \right)$
according to the proof of Proposition 1.
Let ${\cal Y}_1\left(\alpha\right)\triangleq \alpha R_U\left( {\bm \theta}_1 \right) +(1-\alpha){\tilde R}_U^{\star}=\left(R_U\left( {\bm \theta}_1 \right)-{\tilde R}_U^{\star}\right)\alpha+{\tilde R}_U^{\star}$ denote the first term of the minimization function in \eqref{obj_P2}, which linearly decreases with $\alpha$, as $R_U\left( {\bm \theta}_1 \right)-{\tilde R}_U^{\star}< 0$. Moreover, let ${\cal Y}_2\left(\alpha\right)\triangleq \alpha R_C\left( {\bm \theta}_1 \right)$ denote the second term of the minimization function in \eqref{obj_P2}, which linearly increases with $\alpha$. Since we have ${\cal Y}_1\left(0\right)={\tilde R}_U^{\star}>{\cal Y}_2\left(0\right)=0$ and ${\cal Y}_1\left(1\right)=R_U\left( {\bm \theta}_1 \right)<{\cal Y}_2\left(1\right)=R_C\left( {\bm \theta}_1 \right)$, there exists one and only one intersection point in the range of $0\le \alpha \le 1$ such that ${\cal Y}_1\left(\alpha\right)={\cal Y}_2\left(\alpha\right)$, which is the optimal solution to (P2). Thus, by solving ${\cal Y}_1\left(\alpha\right)={\cal Y}_2\left(\alpha\right)$, we obtain the optimal $\alpha^{\star}$ given in \eqref{opt_factor}.
\end{IEEEproof}

\subsubsection{\bf Passive Reflection Optimization} Next, we optimize the passive beamforming ${\bm \theta}_1$ with fixed $\alpha$, for which (P1) is equivalent to\footnote{Note that if we substitute the optimal $\alpha^{\star}$ in \eqref{opt_factor} into \eqref{rate_relay}, the objective function of (P1) turns out to be more complicated with respect to ${\bm \theta}_1$ and thus becomes more difficult to handle. In view of this, we optimize each one of $\alpha$ and ${\bm \theta}_1$ alternately with the other being fixed.}
\begin{align}
\text{(P3):}~
&  \underset{{\bm \theta}_1,\delta}{\text{max}} 
& &  \delta\label{obj_P3}\\
& \text{~~s.t.} & & \alpha R_U\left( {\bm \theta}_1 \right) +(1-\alpha){\tilde R}_U^{\star} \ge \delta,\label{con1_P3}\\ 
& & &\alpha R_C\left( {\bm \theta}_1 \right) \ge \delta,\label{con2_P3}\\
& & & |{\theta_{\mu,1}}|=1, \forall m=1,\ldots,M.\label{con3_P3}
\end{align}
where the constraint of $\rho_C \left( {\bm \theta}_1 \right)> \rho_U \left( {\bm \theta}_1 \right)$ in \eqref{con0_P1} is relaxed without loss of optimality.\footnote{Note that if the constraint $\rho_C \left( {\bm \theta}_1 \right)> \rho_U \left( {\bm \theta}_1 \right)$ does not hold for the obtained ${\bm \theta}_1$ by solving (P3), then we have $C_2^{\star}\ge R_U \left( {\bm \theta}_1 \right)\ge R_C \left( {\bm \theta}_1 \right)\ge \alpha R_C \left( {\bm \theta}_1 \right)\ge C_1\left({\bm \theta}_1,\alpha\right)$ according to \eqref{rate_relay} and \eqref{opt_C2}, which implies that the resultant maximum achievable rate in the relaying IRS case is no greater than that in the conventional IRS case; thus, there is no loss of optimality if we select the maximum rate of these two cases for the considered system.}
After substituting $\rho_U \left( {\bm \theta}_1 \right)$ in \eqref{SNR_U1} and $\rho_C \left( {\bm \theta}_1 \right)$ in  \eqref{SNR_C1} into \eqref{con1_P3} and \eqref{con2_P3}, respectively, and with some simple manipulations, (P3) is equivalently rewritten as
\begin{align}
\hspace{-0.3cm}\text{(P3.1):}~
&  \underset{{\bm \theta}_1,\delta}{\text{max}} 
& &  \delta\label{obj_P3.1}\\
& \text{~~s.t.} & & |h_{\rm AU}+{\bm q}_U^H{\bm \theta}_1|^2\ge \frac{ \sigma^2}{P_A}\left(2^{\frac{\delta}{\alpha} -\frac{1-\alpha}{\alpha}{\tilde R}_U^{\star} }-1\right),\label{con1_P3.1}\\ 
& & &|h_{\rm AC}+{\bm q}_C^H {\bm \theta}_1|^2 \ge \frac{\sigma^2}{P_A}\left(2^{\frac{\delta}{\alpha}}-1\right),\label{con2_P3.1}\\
& & & |{\theta_{\mu,1}}|=1, \forall m=1,\ldots,M.\label{con3_P3.1}
\end{align}
Although the constraints in \eqref{con1_P3.1} and \eqref{con2_P3.1} are still non-convex with respective to ${\bm \theta}_1$, they can be 
equivalently rewritten as
\begin{align}
&{\bar{\bm \theta}}_1^H {\bm B}_U {\bar{\bm \theta}}_1 + |h_{\rm AU}|^2 \ge \frac{ \sigma^2}{P_A}\left(2^{\frac{\delta}{\alpha} -\frac{1-\alpha}{\alpha}{\tilde R}_U^{\star} }-1\right),\label{con1_P3.1eq}\\ 
&{\bar{\bm \theta}}_1^H {\bm B}_C {\bar{\bm \theta}}_1 + |h_{\rm AC}|^2  \ge \frac{\sigma^2}{P_A}\left(2^{\frac{\delta}{\alpha}}-1\right),\label{con2_P3.1eq}
\end{align}
where 
\begin{align}
{\bm B}_U=\begin{bmatrix}
{\bm q}_U {\bm q}_U^H&h_{\rm AU}{\bm q}_U \\h_{\rm AU}^H{\bm q}_U^H,&0
\end{bmatrix},
{\bm B}_C=\begin{bmatrix}
{\bm q}_C {\bm q}_C^H&h_{\rm AC}{\bm q}_C \\h_{\rm AC}^H{\bm q}_C^H,&0
\end{bmatrix},\notag
\end{align}
and ${\bar{\bm \theta}}_1=\left[{\bm \theta}_1^T, t\right]^T$ with $t$ being an auxiliary variable.
As ${\bar{\bm \theta}}_1^H {\bm B}_U{\bar{\bm \theta}}_1=\text{tr} \left({\bm B}_U {\bar{\bm \theta}}_1{\bar{\bm \theta}}_1^H \right)$ and ${\bar{\bm \theta}}_1^H {\bm B}_C{\bar{\bm \theta}}_1=\text{tr} \left({\bm B}_C {\bar{\bm \theta}}_1{\bar{\bm \theta}}_1^H \right)$, we further define ${\bm \Psi}_1={\bar{\bm \theta}}_1{\bar{\bm \theta}}_1^H$, which is required to satisfy ${\bm \Psi}_1 \succeq {\bm 0}$ and $\text{rank}\left({\bm \Psi}_1\right)=1$. Since the rank-one constraint is non-convex, we relax this constraint and transform (P3.1) to
\begin{align}
\hspace{-0.3cm}\text{(P3.2):}~
&  \underset{{\bm \Psi}_1,\delta}{\text{max}} 
& &  \delta\label{obj_P3.2}\\
& \text{~~s.t.} & & \text{tr}\hspace{-0.07cm} \left(\hspace{-0.05cm}{\bm B}_U {\bm \Psi}_1 \hspace{-0.05cm}\right)\hspace{-0.05cm}+\hspace{-0.05cm} |h_{\rm AU}|^2  \hspace{-0.05cm}\ge\hspace{-0.08cm} \frac{ \sigma^2}{P_A}\hspace{-0.08cm}\left(\hspace{-0.05cm}2^{\frac{\delta}{\alpha} \hspace{-0.05cm}-\hspace{-0.05cm}\frac{1-\alpha}{\alpha}{\tilde R}_U^{\star} }-1\hspace{-0.05cm}\right),\hspace{-0.1cm}\label{con1_P3.2}\\ 
& & &\text{tr}\hspace{-0.07cm} \left(\hspace{-0.05cm}{\bm B}_C {\bm \Psi}_1 \hspace{-0.05cm}\right)\hspace{-0.05cm}+\hspace{-0.05cm} |h_{\rm AC}|^2\hspace{-0.05cm} \ge \hspace{-0.08cm}\frac{\sigma^2}{P_A}\hspace{-0.05cm}\left(\hspace{-0.05cm}2^{\frac{\delta}{\alpha}}-1\hspace{-0.05cm}\right),\label{con2_P3.2}\\
& & & \left[{\bm \Psi}_1\right]_{m,m}=1, \forall m=1,\ldots,M+1,\label{con3_P3.2}\\
& & & {\bm \Psi}_1\succeq {\bm 0} .\label{con4_P3.2}
\end{align}
It can be verified that (P3.2) is a quasi-convex optimization problem, which can be efficiently solved by the bisection search: for any given ${\delta}$, (P3.2) reduces to a feasibility-check problem, which is a convex semidefinite program (SDP) and thus can be optimally solved by the existing convex optimization solvers such as CVX \cite{grant2014cvx}.
While the SDR technique may not lead to a rank-one solution, we can retrieve a high-quality rank-one solution to (P3.2) from the obtained higher-rank solution by using e.g., Gaussian randomization \cite{Luo2010Semidefinite}.


In the proposed AO algorithm, we solve (P1) by solving (P2) and (P3.2) alternately in an iterative manner, where the solution obtained in each iteration is used as the initial point for the next iteration. 
The proposed algorithm is guaranteed to converge since the objective value of \eqref{obj_P1} is non-decreasing over the iterations and upper-bounded by a finite value ${\text{min}} \left\{ {\tilde R}_U^{\star},{R}_C^{\star}  \right\}$.
Moreover, the complexity of solving (P3.2) via the SDR and bisection methods is ${\cal O}(M^{4.5} \log (1/\epsilon))$ \cite{Luo2010Semidefinite} with $\epsilon$ being the accuracy of the bisection search, while that of solving (P2) is negligible due to the closed-form solution given in \eqref{opt_factor}.

\section{Simulation Results}\label{Sim} 
\begin{figure}[!t]
	\centering
	\includegraphics[width=2.5in]{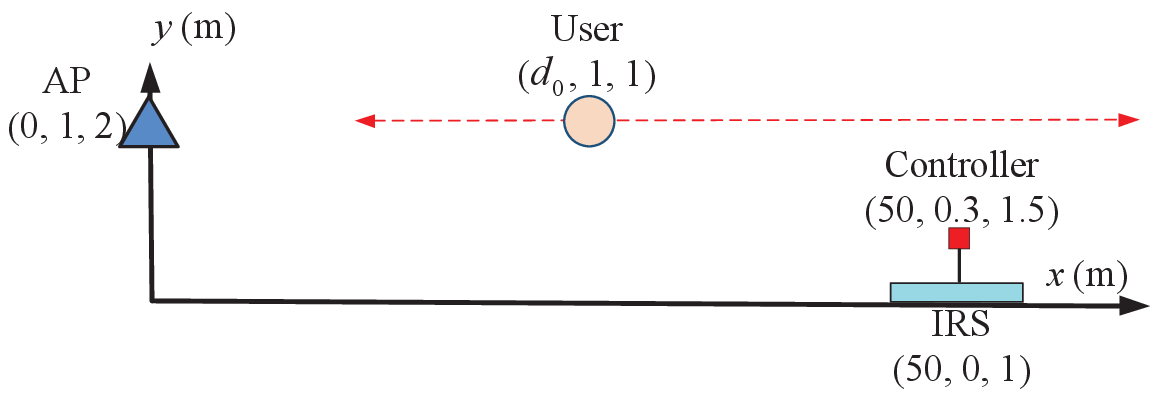}
	\setlength{\abovecaptionskip}{-6pt}
	\caption{Simulation setup (top view).}
	\label{Simulation}
	\vspace{-0.4cm}
\end{figure}

In this section, we present simulation results to examine the performance of the proposed relaying IRS-assisted communication system.
Under a three-dimensional (3D) Cartesian coordinate system, we assume that the AP, IRS, IRS controller, and user are located at
$(0, 1, 2)$, $(50, 0, 1)$, $(50, 0.3, 1.5)$, and $(d_0, 1, 1)$ in meter (m), respectively, as shown in Fig.~\ref{Simulation}, where $d_0$ denotes
the horizontal distance between the AP and the user. For the IRS, we consider the uniform planar array (UPA) consisting of $M=20 \times 20 =400$ elements with half-wavelength spacing placed alone the $x-z$ plane. The distance-dependent channel path loss is modeled as $\gamma=\gamma_0/ d^\alpha$, where $\gamma_0$ denotes the reference path gain at the distance of 1~m which is set as $\gamma_0=-30$~dB for all individual links, $d$ denotes the propagation distance, and $\alpha$ denotes the path loss exponent.
Since the IRS controller and the IRS reflecting elements are co-located with very short distances, the IRS$\rightarrow$controller channel ${\bm h}_{\rm IC}=[h_{{\rm IC}, 1},\ldots, h_{{\rm IC}, M}]$ is modeled by the near-field line-of-sight (LoS) channel with the $m$-th channel coefficient given by $h_{{\rm IC}, m}=\frac{\sqrt{\gamma_0}}{d_{{\rm IC}, m}} e^{-\frac{j 2\pi d_{{\rm IC}, m}}{\lambda}}$, where $d_{{\rm IC}, m}$ is the distance between the $m$-th IRS element and the IRS controller, and $\lambda= 0.05$ m is the wavelength.
Due to the relatively larger distance and random scattering between the AP and the user, the AP$\rightarrow$user channel $h_{\rm AU}$ is characterized by Rayleigh fading with the path loss exponent of $3$; while the remaining channels, i.e., $\left\{{\bm h}_{\rm AI}, h_{\rm AC}, {\bm g}_{\rm IU}, g_{\rm CU}\right\}$ are modeled by Rician fading with the Rician factor of $10$~dB and the path loss exponent of $2.5$.
\rev{The noise power at the IRS controller and the user is set as $\sigma^2=-50$ dBm,
and the transmit power of the AP and the IRS controller is set as $P_A=P_C=8$ dBm so as to keep the same (constant) transmit power in the considered system with or without IRS controller relaying for fair comparison.}

\begin{figure}[!t]
	\centering
	\includegraphics[width=3.5in]{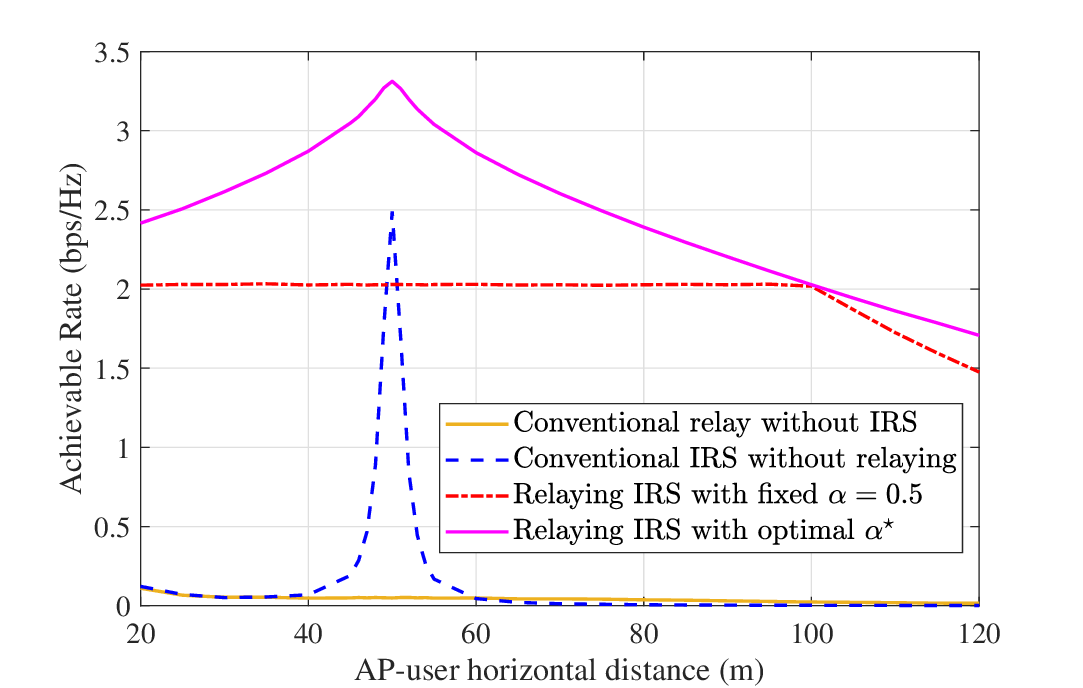}
	\setlength{\abovecaptionskip}{-5pt}
	\caption{Achievable rate versus AP-user horizontal distance $d_0$.}
	\label{rate_SNR_SUz}
	\vspace{-0.4cm}
\end{figure}

In Fig.~\ref{rate_SNR_SUz}, we show the achievable rate versus the AP-user horizontal distance $d_0$ for different cases.
It is observed that by unlocking the IRS controller for opportunistic relaying,
the relaying IRS case with optimal time allocation achieves much better rate performance than the conventional IRS case with passive reflection only, for all values of $d_0$. This is expected since the condition in Proposition~2 usually holds in practice, as in this setup with the co-located IRS controller and IRS reflecting elements. In contrast, the condition in Proposition~1 is much less likely to  occur practically since the IRS-user distance is typically no less than the IRS-controller distance.
Furthermore, even with the equal time allocation $\alpha=0.5$ for the two phases, the relaying IRS case shows an almost flat rate within the range of $20$ m$\le d_0 \le 100$ m, thus substantially enlarging the coverage of the conventional IRS given the target rate of $2$ bps/Hz. Such a flat rate can be explained by the fact that the achievable rate of the relaying IRS case with $\alpha=0.5$ is bottlenecked by the effective channel gain from the AP to IRS controller, which is smaller than that from the IRS controller to user in this distance range. \rev{Finally, it is observed that owing to the large aperture gain brought by the IRS, all the cases aided with IRS significantly outperform the conventional relay (located at the same position as the IRS controller and with the optimized time allocation ratio) without IRS,
especially when the user is located in the vicinity of the IRS/relay (i.e., $40$ m$\le d_0 \le 60$ m).}

\section{Conclusions}\label{conlusion}
In this letter, we proposed a new IRS-assisted communication system by exploiting the IRS controller's relaying capability.
 We jointly optimized the time allocations for DF relaying by the IRS controller and the IRS passive beamforming to maximize the user's achievable rate. Moreover, we showed both analytically and numerically the performance gains of the proposed new design over the conventional IRS with passive signal reflection only, in terms of both user achievable rate and IRS coverage range.

\ifCLASSOPTIONcaptionsoff
  \newpage
\fi

\bibliographystyle{IEEEtran}
\bibliography{RelayIRS}

\begin{thebibliography}{10}
\providecommand{\url}[1]{#1}
\csname url@samestyle\endcsname
\providecommand{\newblock}{\relax}
\providecommand{\bibinfo}[2]{#2}
\providecommand{\BIBentrySTDinterwordspacing}{\spaceskip=0pt\relax}
\providecommand{\BIBentryALTinterwordstretchfactor}{4}
\providecommand{\BIBentryALTinterwordspacing}{\spaceskip=\fontdimen2\font plus
\BIBentryALTinterwordstretchfactor\fontdimen3\font minus
  \fontdimen4\font\relax}
\providecommand{\BIBforeignlanguage}[2]{{%
\expandafter\ifx\csname l@#1\endcsname\relax
\typeout{** WARNING: IEEEtran.bst: No hyphenation pattern has been}%
\typeout{** loaded for the language `#1'. Using the pattern for}%
\typeout{** the default language instead.}%
\else
\language=\csname l@#1\endcsname
\fi
#2}}
\providecommand{\BIBdecl}{\relax}
\BIBdecl

\bibitem{wu2020intelligent}
Q.~Wu, S.~Zhang, B.~Zheng, C.~You, and R.~Zhang, ``Intelligent reflecting
  surface aided wireless communications: A tutorial,'' \emph{IEEE Trans.
  Commun.}, vol.~69, no.~5, pp. 3313--3351, May 2021.

\bibitem{Renzo2020Smart}
M.~{Di Renzo}, A.~{Zappone}, M.~{Debbah}, M.~S. {Alouini}, C.~{Yuen}, J.~{de
  Rosny}, and S.~{Tretyakov}, ``Smart radio environments empowered by
  reconfigurable intelligent surfaces: How it works, state of research, and the
  road ahead,'' \emph{IEEE J. Sel. Areas Commun.}, vol.~38, no.~11, pp.
  2450--2525, Nov. 2020.

\bibitem{Emill2020Intelligent}
E.~{Bj\"{o}rnson}, O.~{\"{O}zdogan}, and E.~G. {Larsson}, ``Intelligent
  reflecting surface versus decode-and-forward: How large surfaces are needed
  to beat relaying?'' \emph{IEEE Wireless Commun. Lett.}, vol.~9, no.~2, pp.
  244--248, Feb. 2020.

\bibitem{Renzo2020Reconfigurable}
M.~{Di Renzo} \emph{et~al.}, ``Reconfigurable intelligent surfaces vs.
  relaying: Differences, similarities, and performance comparison,'' \emph{IEEE
  Open J. Commun. Soc.}, vol.~1, pp. 798--807, Jun. 2020.

\bibitem{Ye2021Spatially}
J.~{Ye}, A.~{Kammoun}, and M.~S. {Alouini}, ``Spatially-distributed {RISs} vs
  relay-assisted systems: A fair comparison,'' \emph{IEEE Open J. Commun.
  Soc.}, doi: 10.1109/OJCOMS.2021.3060929, Feb. 2021.

\bibitem{Yildirim2021Hybrid}
I.~{Yildirim}, F.~{Kilinc}, E.~{Basar}, and G.~C. {Alexandropoulos}, ``Hybrid
  {RIS}-empowered reflection and decode-and-forward relaying for coverage
  extension,'' \emph{IEEE Commun. Lett.}, doi: 10.1109/LCOMM.2021.3054819, Jan.
  2021.

\bibitem{ying2020relay}
X.~Ying, U.~Demirhan, and A.~Alkhateeb, ``Relay aided intelligent
  reconfigurable surfaces: Achieving the potential without so many antennas,''
  \emph{arXiv preprint arXiv:2006.06644}, 2020.

\bibitem{nguyen2021hybrid}
N.~T. Nguyen, J.~He, V.-D. Nguyen, H.~Wymeersch, D.~W.~K. Ng, R.~Schober,
  S.~Chatzinotas, and M.~Juntti, ``Hybrid relay-reflecting intelligent
  surface-aided wireless communications: Opportunities, challenges, and future
  perspectives,'' \emph{arXiv preprint arXiv:2104.02039}, 2021.

\bibitem{Liang2005Gaussian}
Y.~Liang and V.~V. {Veeravalli}, ``Gaussian orthogonal relay channels: Optimal
  resource allocation and capacity,'' \emph{IEEE Trans. Inf. Theory}, vol.~51,
  no.~9, pp. 3284--3289, Sept. 2005.

\bibitem{zheng2019intelligent}
B.~Zheng and R.~Zhang, ``Intelligent reflecting surface-enhanced {OFDM}:
  Channel estimation and reflection optimization,'' \emph{IEEE Wireless Commun.
  Lett.}, vol.~9, no.~4, pp. 518--522, Apr. 2020.

\bibitem{grant2014cvx}
\BIBentryALTinterwordspacing
M.~Grant and S.~Boyd, ``{CVX}: Matlab software for disciplined convex
  programming,'' 2016. [Online]. Available: \url{http://cvxr.com/cvx}
\BIBentrySTDinterwordspacing

\bibitem{Luo2010Semidefinite}
Z.~{Luo}, W.~{Ma}, A.~M. {So}, Y.~{Ye}, and S.~{Zhang}, ``Semidefinite
  relaxation of quadratic optimization problems,'' \emph{IEEE Signal Process.
  Mag.}, vol.~27, no.~3, pp. 20--34, May 2010.

\end{thebibliography}

\end{document}